\documentclass[12pt]{article}%
\usepackage{amsmath}
\usepackage{amsfonts}
\usepackage{amssymb}
\usepackage{graphicx}%
\begin{document}
\begin{center}
\renewcommand{\thefootnote}{\fnsymbol{footnote}}
{\Large\bf Recollections of my work on \lowercase{m}SUGRA with Pran
Nath and Richard Arnowitt\\Circa 1981-85} \vskip20mm {\large\bf{Ali
H. Chamseddine \footnote{Published in "Themes in Unification: The
Pran Nath Festschrift"
PASCOS 2004, Editors George Alverson and Michael Vaughn, World Scientific 2005.}}}\\
\renewcommand{\thefootnote}{\arabic{footnote}}
\vskip2cm {\it Center for Advanced Mathematical Sciences (CAMS)
and\\
Physics Department, American University of Beirut, Lebanon\\
Email: chams@aub.edu.lb}
\end{center}
\vspace{1cm}
 In the early 1975 when I\ was a graduate student of
Abdus Salam at Imperial College, London, working on the topic of
supersymmetry, Salam gave me a paper he had recently received from
Dick Arnowitt and Pran Nath where the new idea of supergravity in
superspace was introduced\cite{superspace}. Salam asked me to study
this paper and work on this topic if I found it interesting. I
thought that this was a brilliant idea, and the topic became part of
my thesis\cite{thesis}. This was my first encounter with Pran and
Dick although it was only through their work. It turned out that
this was only a prelude to an extended interaction at the
collaborative level that was soon to come. Thus in the Fall of 1980
I\ was a scientific associate at CERN when I\ met Pran there as he
was spending his sabbatical leave at CERN. Soon after our first
meeting he made me an offer to visit Northeastern University as a
research associate for one year. At that time Lebanon was engulfed
in a savage civil war, so I readily accepted his kind offer as it
gave me and my family a safe shelter and a good environment to
pursue my research. In January 1981 I\ joined Northeastern
University, and at first I\ continued my work on $N=1$
ten-dimensional supergravity\cite{10d} \ and I succeeded in
constructing the coupling of supergravity to Yang-Mills gauge
supersymmetry, and compactifying the system to
four-dimensions\cite{10dYM}. The problem I\ faced was that the
compactified four-dimensional system has only $N=4,$ $N=2$ or $N=0$
supersymmetry and it appeared difficult to get an $N=1$ compactified
theory since at that time the Calabi-Yau compactification was not
familiar to physicists\cite{Candelas}. On the other hand unified
models based on $N=1$ supersymmetry appeared to resolve the problem
of gauge hierarchy at least at the technical level\cite{georgi} and
hence it was imperative that further work on model building utilize
the framework of $N=1$ supersymmetry.

After the analysis on the reduction of ten-dimensional supergravity was
completed, I started interacting with Pran and Dick. At that time they were
working on supergravity in superspace and on the U(1) axial anomaly
\cite{U(1)}. However, as a consequence of our interactions our interests
converged on model building based on $N=1$ supergravity. This was a novel idea
as there were no phenomenologically viable models of this type in the
literature at that time. My discussions with Pran and Dick started in earnest
in September 1981 on ways of obtaining realistic $N=1$ supergravity
interacting with $N=1$ super Yang-Mills and chiral $N=1$ multiplets. The first
attempt was to realize such a construction from ten-dimensions. However, as
explained already, at that time there was no known way of obtaining $N=1$
supersymmetry from higher dimensions. Thus we decided to construct the general
$N=1$ supergravity interactions directly in four-dimensions. A Lagrangian with
the most general coupling of one chiral multiplet had already been constructed
in 1979 by Cremmer et al\cite{cremmer79} using the methods of superconformal
tensor calculus\cite{superconformal}. We used this method to construct the
general $N=1$ supergravity Lagrangian coupled to super Yang-Mills multiplets
and an arbitrary number of chiral multiplets. This proved to be a rather
complicated task which we were able to finish in early spring of 1982. The
results were very interesting, and the form of the supergravity scalar
potential which contained both positive and negative contributions implied
that it was possible to break supersymmetry spontaneously and obtain a zero
cosmological constant, which is essential to obtain a realistic model. The
most general interactions involved an arbitrary function of the scalar fields,
denoted by $\mathcal{G}$ which can be split into a Kahler part and a
superpotential part which was dictated by requiring that when the limit
$M\left(  \text{Planck}\right)  \rightarrow\infty$ is taken the action reduces
to that of global supersymmetry.

Although we had all the results on the $N=1$ applied supergravity in early
spring of 1982 we did not immediately publish them (they were later published
in Trieste Lectures Series\cite{book}) since there were some other weighty
ideas we were after and these included the construction of a realistic model
of particle interactions within the $N=1$ supergravity framework where
supersymmetry was broken by a super Higgs mechanism. The main aim was to
obtain soft breaking including mass growth for the sparticles which overcame
the pitfalls of models based on global supersymmetry where, for example,
spontaneous breaking of supersymmetry leads to a squark having mass less than
that of a quark. Thus beginning in early Spring of 1982, our efforts over the
next few months were focused in this direction. There were several hurdles to
be overcome. The first was to break supersymmetry and adjust the vacuum energy
to zero. This could be done by breaking supersymmetry by a super Higgs
mechanism, and utilizing the fact that the scalar potential of the model was
not positive definite, to fix the vacuum energy to zero. The second was to
protect the low energy theory below the Planck scale from mass growth of the
size of the Planck mass. Such a mass growth would arise naturally if the super
Higgs mechanism occurred in the same sector where the quarks, leptons and
other matter fields reside in the superpotential. To overcome this hurdle the
superpotential was split into two different sectors, a (visible) sector where
visible matter, i.e., quarks, leptons, and Higgs, reside and a (hidden) sector
where the super Higgs mechanism operates. The key idea here was to have no
direct interaction between these two sectors. Because of a lack of this direct
interaction soft masses in the physical sector of the size of the Planck scale
are avoided. On the other hand, the two sectors are coupled by gravitational
interactions because of the supergravity structure of the scalar potential. An
interesting question then arises, what is the implication of breaking of
supersymmetry in the hidden sector on the visible sector?. We addressed this
issue by deducing the effective low energy theory in the visible sector. The
result of the analysis was very interesting in that the scalar fields in the
visible sector showed mass growths of size $O(m^{2}/M_{Planck})$ where $m$ is
an effective intermediate scale that appears in the super Higgs effect. Thus
with $m\sim10^{10}$ GeV, soft masses of size $O(10^{2-3})$ GeV could be
generated. Additionally we found that there were soft bilinear and trilinear
couplings in the effective theory before the Planck scale. The nature of soft
breaking depends on the nature of Kahler potential chosen and for the analysis
we performed the Kahler potential was assumed to be flat. Consequently our
analysis exhibited a universality of the soft parameters.

There are two further phenomena which need to be commented on in this initial
work on supergravity model building. The first one is that the model we were
working with was a grand unified supergravity model. And in this model the
breaking of supersymmetry and of grand unification was accomplished in one
step. Quite remarkably the soft breaking was found to be independent of not
just the Planck scale but also of the grand unification scale $M_{G}$. Second,
in our analysis we showed that the soft breaking lead to the breaking of the
electroweak symmetry from $SU(2)_{L}\times U(1)_{Y}$ to $U(1)_{em}$. Thus
together these phenomena produced a supergravity grand unification with soft
breaking of electroweak size and provided also for an explanation for the
breaking of the electroweak symmetry. All these results are contained in our
first paper Ref.\cite{locally}. After the submission of our work to Physical
Review Letters\cite{locally} we became aware of the work of Cremmer et al on
the couplings of $N=1$ supergravity\cite{cremmer82}. However, this paper did
not contain formulation of a SUGRA model with hidden sector breaking.
Immediately after, we received  the work of Ref\cite{barbieri} which also
achieved soft breaking of supersymmetry through the hidden sector mechanism.
However, this work did not contain a grand unification nor an exhibition of
the phenomenon that the low energy theory was independent of $M_{G}$.

I\ spent the next three years working hard in a very fruitful collaboration
with Dick and Pran on different applications of $N=1$ supergravity pushing the
idea to its limits which resulted in many interesting additional
works\cite{1,2,3,4,5,6,7,8,9}. I\ learned from them how to work as part of a
team, spending endless hours in discussions. They served for me an excellent
example of how to dedicate oneself to science. With Pran I\ found the friend
who was eager to help long after I left Northeastern. I\ am glad for having
his friendship all these years.

\end{document}